\documentclass[a4paper,12pt]{article}

\usepackage{monash}
\bibliography{refs}

\usepackage{color}
\definecolor{keyword}{HTML}{008000}
\definecolor{emph}{HTML}{0000FF}
\definecolor{string}{HTML}{A52A2A}
\definecolor{comment}{HTML}{004461}
\definecolor{back}{HTML}{F8F8F8}
\definecolor{arrow}{HTML}{745334}

\usepackage{listings}

\newcommand{\superplot}{\code{Superplot}\xspace}
\newcommand{\multinest}{\code{MultiNest}\xspace}

\newcommand{\convolve}{\ast}
\newcommand{\dtft}[1]{\mathcal{F}\left\{#1\right\}}
\newcommand{\idtft}[1]{\mathcal{F}^{-1}\left\{#1\right\}}

\newcommand{\footurl}[1]{\footnote{See \url{#1}.}}

\newcommand{\dif}[1]{\,{d}#1}

\newcommand{\like}{\mathcal{L}}

\newcommand{\home}{\raisebox{-0.5ex}{\textasciitilde}}

\begin{document}

\title{\superplot: a graphical interface for plotting and analysing \multinest output}
\date{\today}
\author{Andrew Fowlie}
\affiliation{ARC Centre of Excellence for Particle Physics at the Tera-scale, School of Physics and Astronomy, Monash University, Melbourne, Victoria 3800 Australia}
\author{Michael Hugh Bardsley}
\affiliation{School of Physics and Astronomy, Monash University, Melbourne, Victoria 3800 Australia}

\date{\today}

\maketitle

%%%%%%%%%%%%%%%%%%%%%%%%%%%%%%%%%%%%%%%%%%%%%%%%%%%%%%%%%%%%%%%%%%%%%%%%%%%%%%%%

\begin{abstract}
We present an application, \superplot, for calculating and plotting statistical quantities relevant to parameter inference
from a ``chain'' of samples drawn from a parameter space, produced by \eg \multinest. 
A simple graphical interface
allows one to browse a chain of many variables quickly, and make publication quality plots
of, inter alia, one- and two-dimensional profile likelihood, posterior pdf (with kernel density estimation), confidence intervals and credible regions. 
In this short manual, we document installation and basic usage, and define all statistical quantities and conventions.
The code is fully compatible with Linux, Windows and Mac OSX.
Furthermore, if preferred, all functionality is available through the command line rather than a graphical interface.
\end{abstract}

%%%%%%%%%%%%%%%%%%%%%%%%%%%%%%%%%%%%%%%%%%%%%%%%%%%%%%%%%%%%%%%%%%%%%%%%%%%%%%%%

\section{Introduction}

Many branches of physics are utilising sophisticated numerical methods to infer the parameters of a model from data. This typically involves a numerical exploration of a parameter space with a Monte-Carlo algorithm, resulting in a collection of weighted samples drawn from the parameter space (henceforth referred to as a chain). A modern example of such an algorithm is nested sampling\cite{Skilling:2004,skilling2006}. The popular \multinest\cite{Feroz:2007kg,Feroz:2008xx,2013arXiv1306.2144F} implementation of nested sampling is utilised for parameter extraction in manifold areas of physics, including supersymmetric fits \see{Fowlie:2014xha,deAustri:2006jwj,sbweb,deVries:2015hva}, cosmological fits \see{Lewis:2002ah,Lewis:2013hha,Zuntz:2014csq,Aslanyan:2013opa,Mortonson:2010er,2012PhRvD..85j3533E,Norena:2012rs} and X-ray analysis \see{Olamaie:2013vfa,Buchner:2014nha}, and in a forthcoming general purpose fitting program, \code{GAMBIT}\cite{gambit}.

We present an application, \superplot, for plotting parameters extracted by \multinest (or a similar code, such as \code{PolyChord}\cite{Handley:2015fda,2015MNRAS.453.4384H}) with the \code{matplotlib} plotting library\cite{matplotlib}. This should simplify the final step in parameter extraction: calculating and plotting results from a chain, such as posterior density, profile likelihood, credible regions or confidence intervals. This is, of course, already possible with private scripts or \code{pippi}\cite{Scott:2012qh}, \code{SuperEGO}\cite{superego}/\code{CosmoloGUI}\cite{cosmologui}, modified versions of the programs \code{GetPlots}\cite{getplots}/\code{GetDist}\cite{getdist} and even \code{ROOT}\cite{Brun:1997pa}. The advantage of \superplot is that a graphical interface allows one to quickly browse a chain of many variables and make publication quality plots, with control over binning or kernel density estimation (KDE), without writing any scripts or codes. We describe installation instructions in \refsec{sec:install} and usage in \refsec{sec:use}. Definitions and conventions of all statistical quantities are provided in \refapp{app:stat}.

% neutral networks \code{BAMBI}\cite{2014MNRAS.441.1741G,2012MNRAS.421..169G}
% galaxy spectral energy distributions \code{BayesSED}\cite{Olamaie:2013vfa}
% gravitational lensing \code{Lensed}\cite{2015arXiv150507674T}
% supernovae \code{pSNid}\cite{2011ApJ...738..162S}
% optical interferometry \code{simtoi}
% pulsar timing \code{TempNest}

\section{Installation}\label{sec:install}

\superplot is a Python 2.7 code. The simplest method of installing \superplot is via the \code{pip} package manager.\footnote{\code{pip} is included in Python beginning in version 2.7.9. If your version of Python does ­­­not include \code{pip}, see \url{https://pip.pypa.io/en/latest/installing/}}
The command
\begin{lstlisting}
pip install superplot
\end{lstlisting}
should install \superplot and dependencies, except \code{matplotlib}. Note, however, that on some operating systems, \code{pip} may not be able to automatically build dependencies. Thus \code{matplotlib} and other dependencies may have to be separately installed; see \refsec{sec:required}. 

\superplot was tested on Linux, Windows and Mac OSX. All plotting functionality and summary statistics are available from the command line on Linux, Windows and Mac OSX, should you experience issues with the graphical interface on your system (see \refsec{Sec:CommandLine}). In Linux, programs are placed in a platform-dependent directory, \eg \code{\home/.local/bin} for Ubuntu and Mint. To invoke the programs directly, \eg \code{superplot\_gui}, this directory must be in the user's \code{path}.

If installation via \code{pip} is problematic, clone the source code from git-hub:
\begin{lstlisting}
git clone https://github.com/michaelhb/superplot.git
\end{lstlisting}
or download the source code from \url{https://github.com/michaelhb/superplot/archive/master.zip}. You may need to refer to \refsec{sec:required} for instructions on installing dependencies. Once all dependencies are satisfied, \superplot can be installed and run from the source directory:
\begin{lstlisting}
python setup.py install
python ./superplot/super_gui.py
\end{lstlisting}

Finally, you may wish to place configuration and example files in a convenient location,
\begin{lstlisting}
superplot_create_home_dir -d <path_to_directory>  # Linux (if ~/.local/bin in path) or
python -m superplot.create_home_dir -d <path_to_directory>  # Linux/Windows/Mac OSX
\end{lstlisting}
See \refsec{sec:custom} for further details.

\subsection{Dependencies}\label{sec:required}
\superplot requires some common Python modules. The most obscure dependencies can be installed via \code{pip}:
\begin{lstlisting}
pip install appdirs prettytable simpleyaml joblib
\end{lstlisting}
but should be installed automatically by \code{pip install superplot}. Others dependencies may require OS-specific installation, including \code{matplotlib}\cite{matplotlib} version 1.4 or newer with \code{gtk} support, \code{numpy}\cite{numpy}, \code{scipy}\cite{scipy} and \code{pandas}\cite{pandas} from the SciPy stack\footurl{http://www.scipy.org/install.html}, and \code{PyGTK}.\footurl{http://www.pygtk.org/downloads.html} For \eg Ubuntu 16 users, dependencies may be installed via
\begin{lstlisting}
apt-get install git python-pip python-numpy python-scipy python-pandas libfreetype6-dev python-gtk2-dev python-matplotlib
\end{lstlisting}
For users of other Linux distributions, including Mint, it may be neccessary to install missing dependencies \code{python-setuptools}, \code{python-tk} and \code{python-wheel} via \eg
\begin{lstlisting}
apt-get install python-setuptools python-tk python-wheel
\end{lstlisting}
Whereas for Windows users, dependencies may be installed by
\begin{enumerate}
\item Installing the Anaconda Python distribution from \url{https://www.continuum.io/downloads}
\item Installing the \code{PyGTK} all-in-one bundle from \url{http://ftp.acc.umu.se/pub/GNOME/binaries/win32/pygtk/2.24}
\item Upgrading \code{matplotlib}, \code{pip install --force-reinstall --no-deps --upgrade matplotlib}
\item Finally, installing \superplot, \code{pip install superplot}
\end{enumerate}
Finally, Mac OSX users must install \code{pygtk} and \code{matplotlib} with \code{pygtk} support, \eg via \code{brew}\footnote{In the past there were issues with \code{matplotlib} and \code{pygtk} in Mac OSX; we believe that they are now resolved.}
\begin{lstlisting}
brew install pygtk
brew install homebrew/python/matplotlib --with-pygtk
\end{lstlisting}
\section{Quickstart}\label{sec:use}

\begin{figure}[h]
\centering
\subfloat[t][First, open a \code{*.txt} file from \multinest.]{\label{fig:gui_txt}
\includegraphics[width=0.35\textwidth]{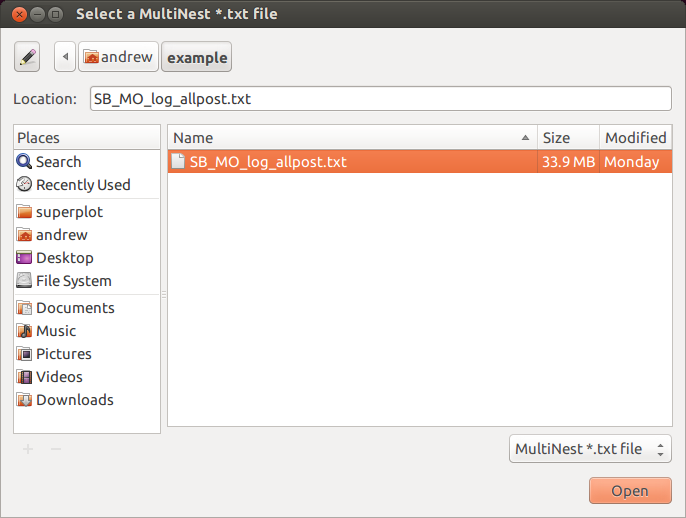}
}%
\subfloat[t][Second, optionally, open a \code{*.info} file that labels columns in the \code{*.txt} file.]{\label{fig:gui_info}
\includegraphics[width=0.35\textwidth]{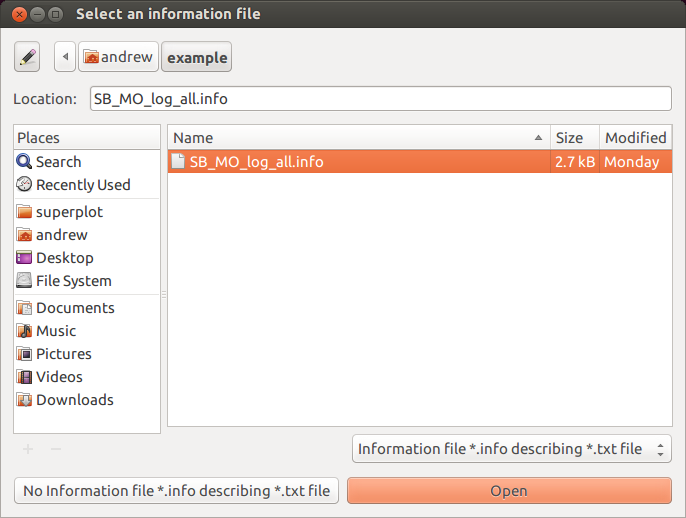}
}%

\subfloat[t][Finally, select a plot, and click \code{Make plot.} in the graphical interface in \superplot.]{\label{fig:gui}
\includegraphics[width=0.7\textwidth]{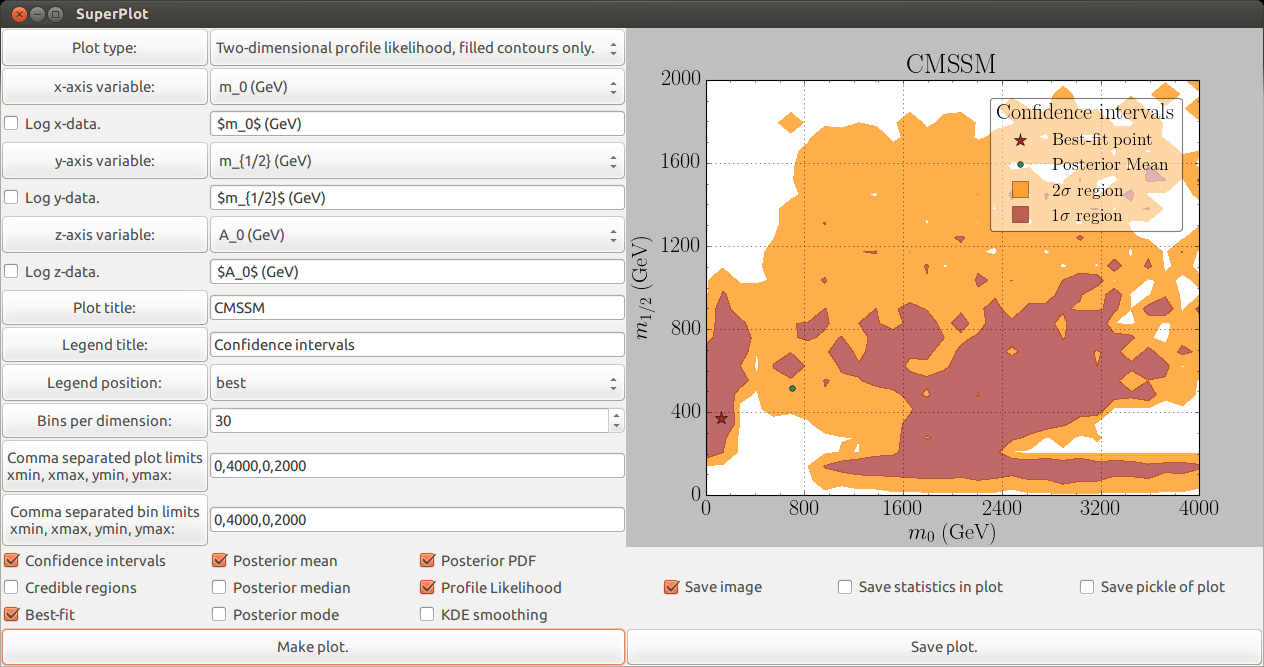}
}%
\caption{Selecting data files for \superplot.}
\label{fig:gui_examples}
\end{figure}

The main component of \superplot is \code{superplot\_gui} --- a graphical interface for making plots from a chain. To start \code{superplot\_gui}, from any directory run either:
\begin{lstlisting}
superplot_gui  # Linux (if ~/.local/bin in path) or
python -m superplot.super_gui  # Linux/Windows/Mac OSX
\end{lstlisting}
The latter is advised for Windows as it may avoid problems with \code{stdout}. You will be prompted to select a \multinest chain ending in \code{*.txt} (\reffig{fig:gui_txt}). Select a chain of your choice \eg \code{SB\_MO\_log\_allpost.txt} --- a chain from \code{SuperBayeS}\cite{deAustri:2006jwj,sbweb}, see \refsec{sec:custom} for further details about its location --- and click \code{Open}. You will be asked whether you wish to select an information file (\reffig{fig:gui_info}). This optional file could contain labels and metadata associated with the chain (see \refsec{sec:info}). Select \eg \code{SB\_MO\_log\_all.info} and click \code{Open}. If you select \code{No information file...}, the variables in the chain will be assigned numerical labels based on column order.

\begin{figure}[ht]
\centering
\subfloat[One-dimensional plot]{\label{fig:1D}
\includegraphics[scale=0.3]{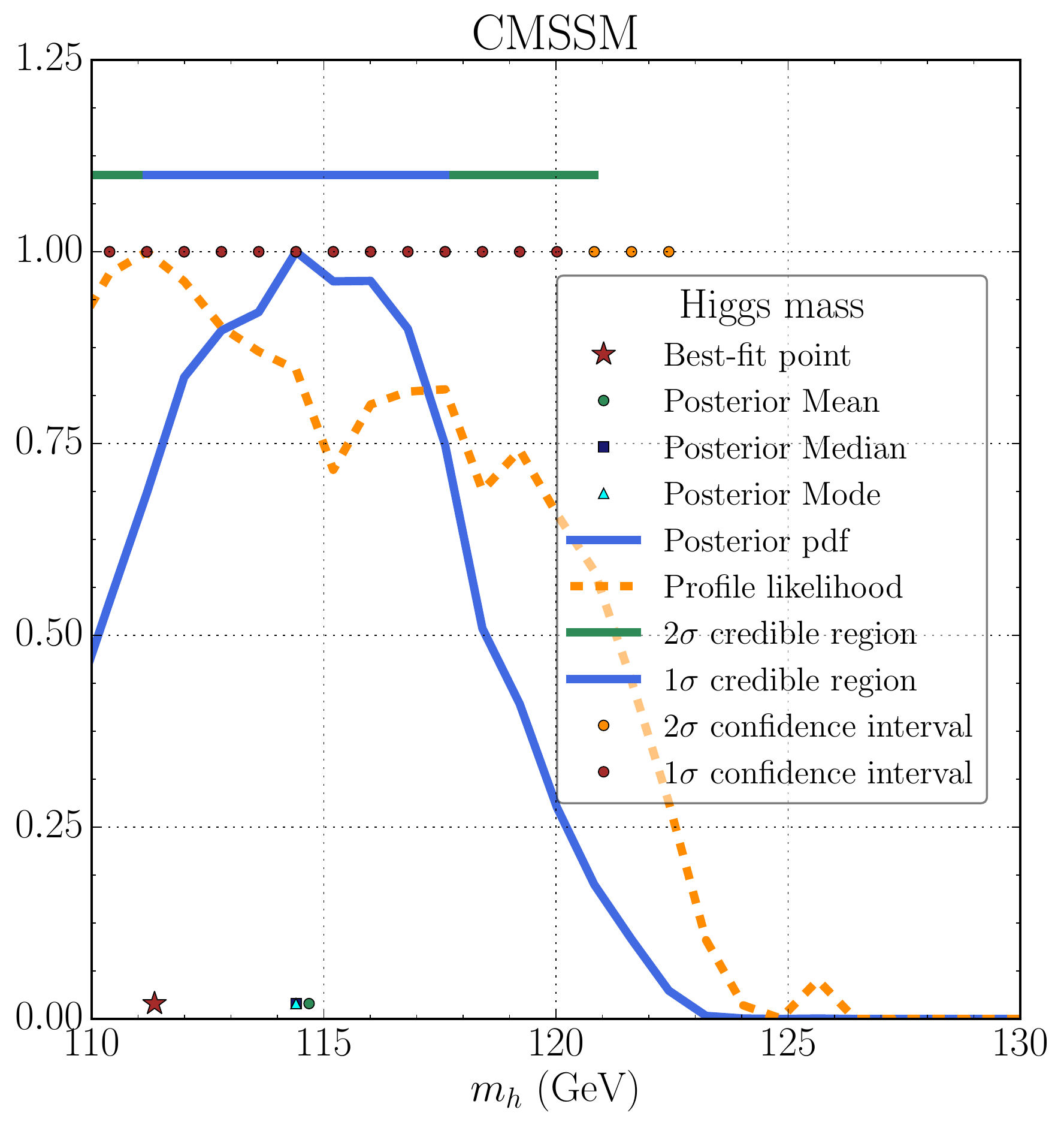}
}%
\hspace*{4.5mm}%
\subfloat[Two-dimensional posterior pdf, filled contours only]{\label{fig:2D}
\includegraphics[scale=0.3]{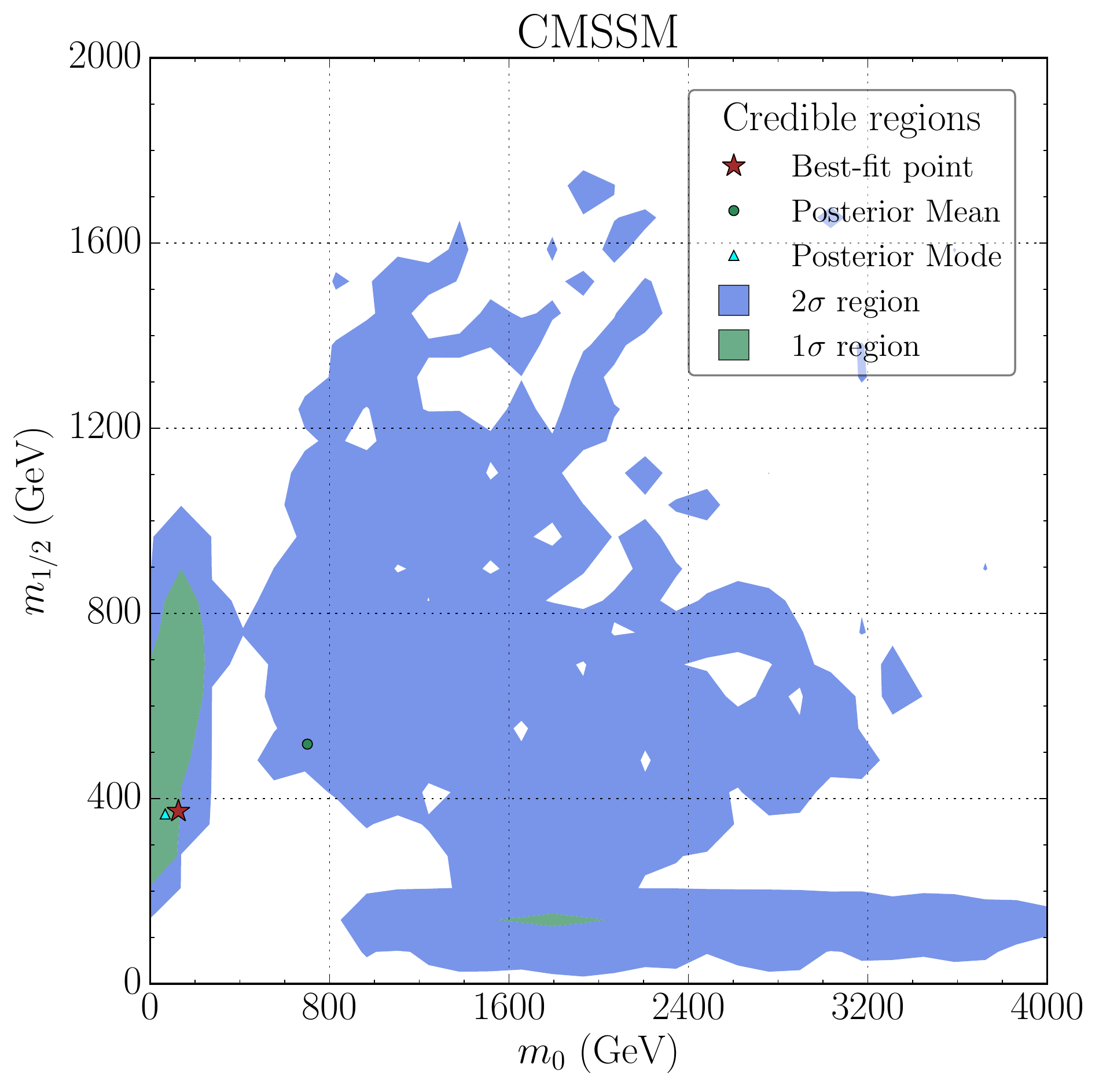}
}%

\subfloat[Two-dimensional profile likelihood, filled contours only]{\label{fig:2D_like}
\includegraphics[scale=0.3]{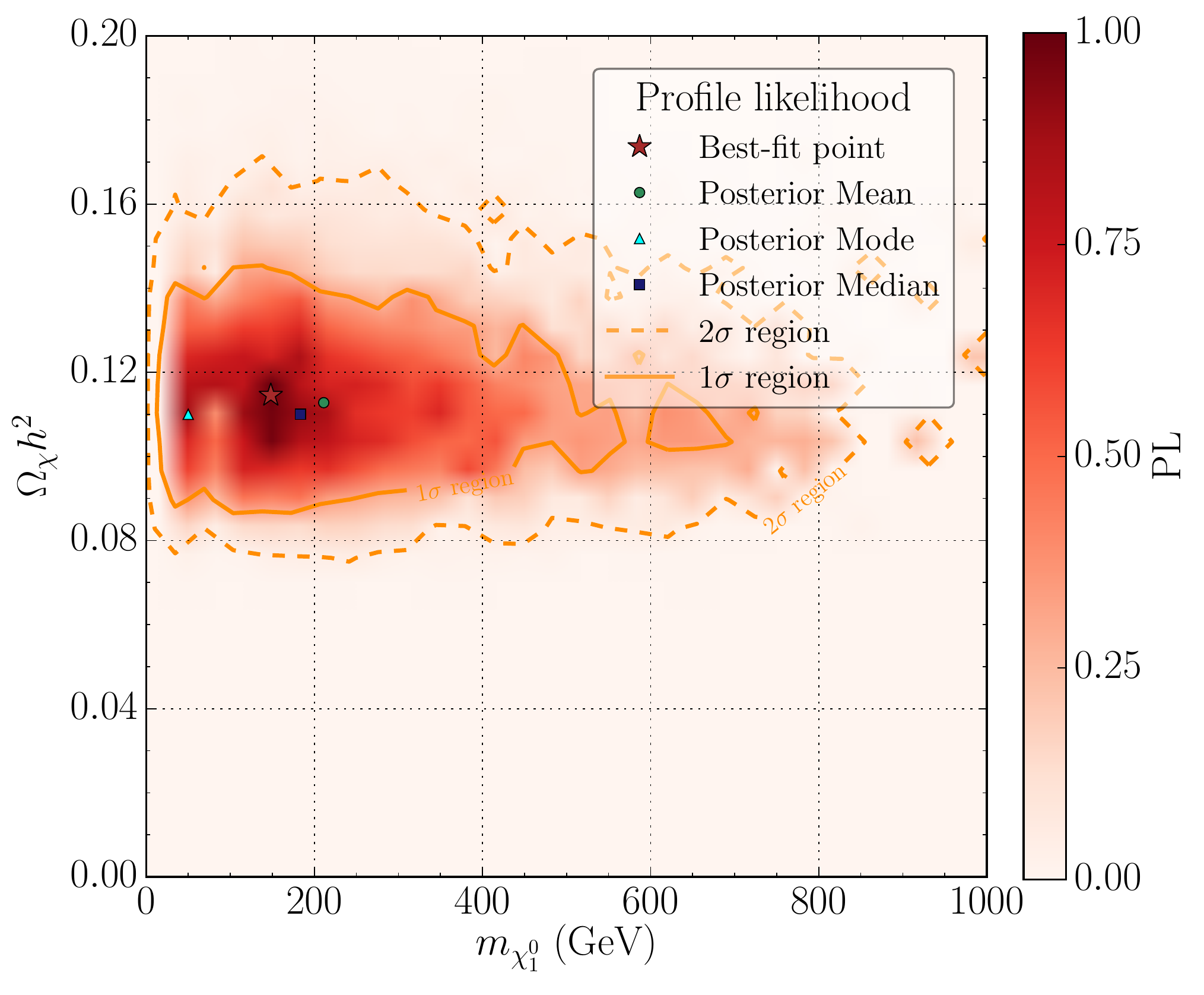}
}%
\subfloat[Two-dimensional posterior pdf]{\label{fig:2D_density}
\includegraphics[scale=0.3]{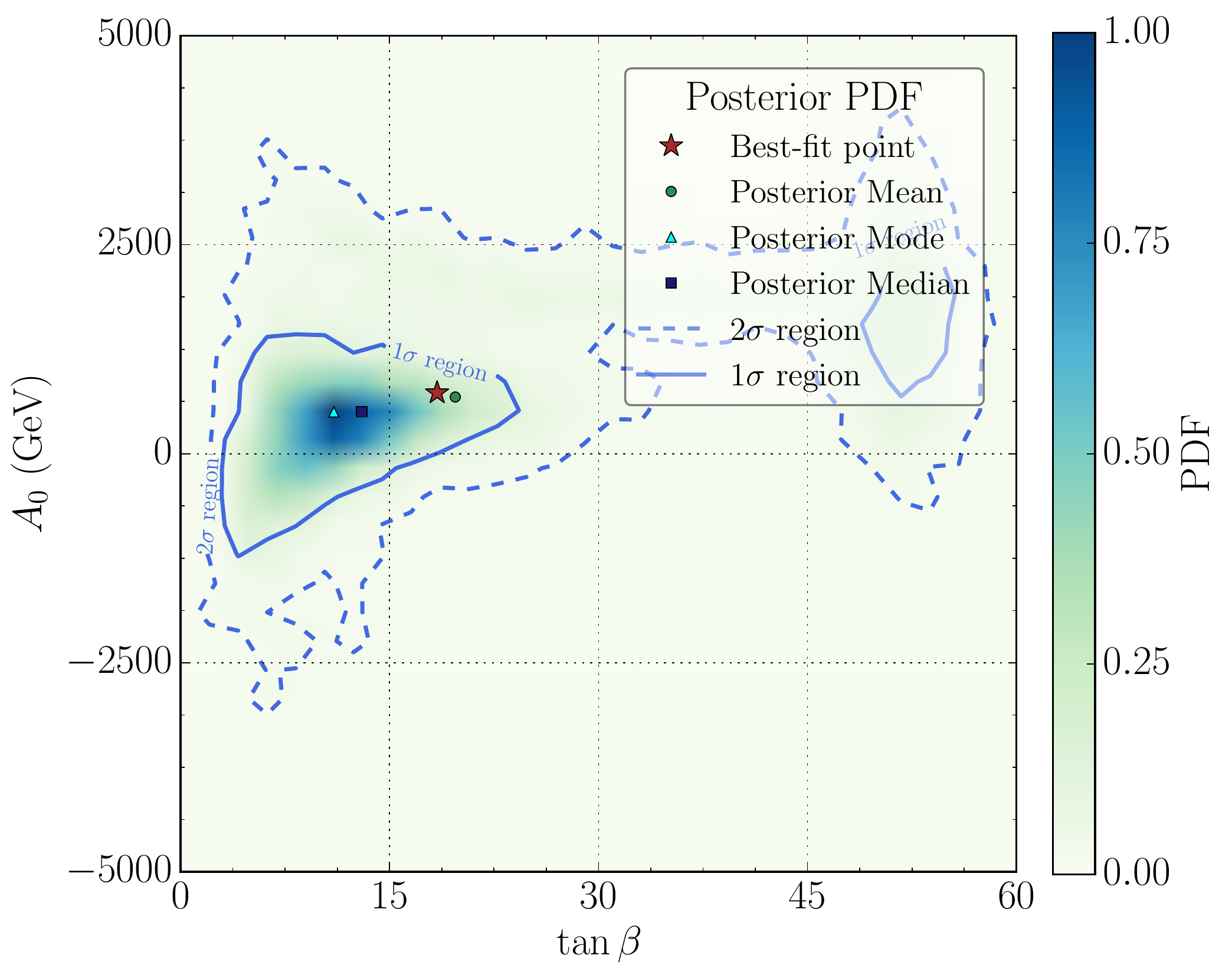}
}%
\caption{Examples of figures produced by \superplot from a publicly released chain from \code{SuperBayeS}\cite{deAustri:2006jwj,sbweb}.}
\label{fig:examples}
\end{figure}

After the information file is selected, the main graphical interface appears (\reffig{fig:gui}). The left-hand side of the window contains controls for configuring the plot, including:
\begin{itemize}
\item Type of plot (\code{Plot type}); see \reffig{fig:examples} for examples. The possibilities are:
\begin{description}
\item \code{One-dimensional plot} One-dimensional marginalised pdf, $p(x)$, and/or profile likelihood, $\like(x)$.
\item \code{One-dimensional chi-squared plot} One-dimensional chi-squared with an theoretical error band.
\item \code{Two-dimensional posterior pdf, filled contours only} Two-dimensional credible regions of posterior pdf, illustrated by filled contours.
\item \code{Two-dimensional profile likelihood, filled contours only} Two-dimensional confidence intervals, illustrated by filled contours.
\item \code{Two-dimensional posterior pdf} Two-dimensional posterior pdf, $p(x,y)$, illustrated by shading on a two-dimensional plane.
\item \code{Two-dimensional profile likelihood} Two-dimensional profile likelihood, $\like(x,y)$, illustrated by shading on a two-dimensional plane.
\item \code{Three-dimensional scatter plot} Three-dimensional scatter plot --- all samples scattered on a two-dimensional plane, shaded by the value of a third variable.
\end{description}

\item Variables you wish to plot (\eg \code{x-axis variable}).
\item The axis labels \eg type the $x$-axis label in text-box below \code{x-axis variable}. Labels may include a \LaTeX\ math environment (\code{\$...\$}) and are, by default, rendered with \code{pdflatex}. Any \code{pdflatex} errors should be printed to the terminal. 
\item Whether a variable should be logged (\eg \code{Log x-data}). If selected, data is logged and then binned. This differs from binning and plotting on a logarithmic scale.
\item The number of bins per dimension and bin limits.
\item The limits for the $x$- and $y$-axis.
\item Plot title, legend title and legend position.
\item Selection of optional plot elements, \eg the best-fit point or posterior mean.
\item Whether to use kernel density estimation (KDE) of pdfs, as described in \refapp{app:KDE}.
\end{itemize}
Once you have selected the plot you wish, click the \code{Make plot} button, located below the plot options. The desired plot should appear in the central window, as in \reffig{fig:gui}. If no plot appears, check whether any \eg \LaTeX\ errors were printed to the terminal. If so, fix any malformed \LaTeX\ and click \code{Make plot}.

The controls for saving plots are located below the central window. The option \code{Save image} creates a high-quality PDF version of the displayed plot, \code{Save statistics in plot} writes a text file containing summary statistics and metadata about the plot, and \code{Save pickle of plot} writes a serialised copy of the \code{matplotlib} plot object.\footnote{See \url{http://matplotlib.org/users/whats_new.html\#figures-are-picklable}. A pickle is a serialised copy of the plot object which can be loaded in a separate Python session. This may be useful if a plot needs to be customised in a manner not otherwise possible. For an example, see \code{example/load\_pkl.py}.} After selecting the desired outputs, press the \code{Save plot} button and select a location on disk.

\subsection{Configuration and example files}\label{sec:custom}

Beyond the control panel in the graphical interface, the appearance of a plot can be customised by editing configuration files included in \superplot. If \superplot was installed via \code{pip}, configuration and example files are located in platform-dependent user data directory. To place them in a directory of your choice, 
\begin{lstlisting}
superplot_create_home_dir -d <path_to_directory>  # Linux (if ~/.local/bin in path) or
python -m superplot.create_home_dir -d <path_to_directory>  # Linux/Windows/Mac OSX
\end{lstlisting}
These files are used in preference to any copies installed with the source code. If they are deleted, \superplot will revert to copies installed with the source code. If the script is run more than once, \superplot will use the configuration files in the most recently created directory. 

If the source code was downloaded, configuration files are included in \code{superplot/config.yml} and \code{superplot/plotlib/styles}, and example files are included in \code{superplot/example}. For further documentation of the code itself, see the API.\footurl{http://superplot.readthedocs.org}

The \code{yaml} file \code{config.yml} contains ``schemes'' which control the sizes, colours, symbols and labelling of individual plot elements, \eg the symbol for the best-fit point. These are specified using \code{matplotlib} conventions --- see the file header for further information. The \code{styles/} directory contains a set of \code{matplotlib} style sheets.\footurl{http://matplotlib.org/users/style_sheets.html} Options such as line thickness, grid lines and fonts can be fine-tuned by editing these files. The file \code{default.mplstyle} contains options which apply to all plots. There are also style sheets specific to each plot type. The individual style sheets take precedence over \code{default.mplstyle}, allowing plot-specific customisation. The  \code{example/} directory contains, inter alia, \code{*.txt} and \code{*.info} example files for \superplot.

\subsection{Summary statistics for chain}

\superplot also includes a command that generates a table of statistics for a chain. This can be launched by running either:
\begin{lstlisting}
superplot_summary --data_file DATA_FILE [--info_file INFO_FILE]  # Linux (if ~/.local/bin in path) or
python -m superplot.summary --data_file DATA_FILE [--info_file INFO_FILE]  # Linux/Windows/Mac OSX
\end{lstlisting}
You should specify a chain and an (optional) information file by the command-line arguments. A table containing the label, best-fit point, posterior mean and $1\sigma$ credible region for every variable in the chain is printed to the terminal. 

\subsection{Making \superplot plot via the command line}\label{Sec:CommandLine}

It is possible to make a \superplot plot from the command line, bypassing the GUI. For all possible options and usage, see
\begin{lstlisting}
python -m superplot.super_command --help  # Linux/Windows/Mac OSX
\end{lstlisting}
The command-line interface may be necessary on systems that fail \code{PyGTK} dependencies. Examples of simple and more complicated usage are
\begin{lstlisting}
# Produce a 1D plot with many default options 
# You may need to alter the path for the *.txt file
python -m superplot.super_command ~/.local/share/superplot/example/SB_MO_log_allpost.txt --xindex=4 
# Produce a 2D plot, with many options specified at the command line
# You may need to alter the path for the *.txt file
python -m superplot.super_command ~/.local/share/superplot/example/SB_MO_log_allpost.txt --xindex=4 --yindex=5 --xlabel='$x$-label' --ylabel='$y$-label' --logy=True --kde=True --plot_title='Example plot' --output_file='example.pdf' 
\end{lstlisting}
The code saves a \superplot plot to disk. In the first case, a descriptive name for the plot with \code{.pdf} extension is chosen by the code and printed to the screen. In the second case, the file name and extension \code{example.pdf} are specified at the command line. There is one compulsory positional argument --- the name of the \code{*.txt} file --- and one compulsory named argument --- the index of the variable on the $x$-axis, \code{--xindex=}. All other arguments are optional and are specified explicitly by name. 

The command line accepts a full set of plot options (including options otherwise only specified in the \code{yaml} file \code{config.yml}), and invokes the same plotting and statistical libraries as the GUI. 

\subsection{Use with programs other than \multinest}

\superplot reads data from the \multinest text file (\code{*.txt}), which is a plain-text array of floats separated by at least one space. The first and second columns are the posterior weight and $-2\ln\like$ or chi-squared of each sample, respectively. Subsequent columns are parameter values associated with each sample. \superplot would work with any data in this format. If you consider only frequentist statistics and calculate only a chi-squared for each sample, the first column could be a place-holder --- the frequentist quantities could still be plotted.

\subsection{Information file}\label{sec:info}

We inherited the format of an information file from \code{getdist}. The information file (\code{*.info}) provides labels for the parameters in the chain, beginning with the third column (\ie ignoring posterior weight and $-2\ln\like$). The \code{*.info} file is optional. If provided, \superplot automatically labels columns of the chain with the \code{*.info} file. The format of an entry in the \code{*.info} file is \eg
\begin{lstlisting}
lab1 = $m_0$ (GeV)
\end{lstlisting}
This would label the first parameter (\ie third column) ``$m_0$ (GeV)''. You do not have to provide labels for all columns. The label may include a \LaTeX\ math environment (\code{\$...\$}) and should not be enclosed in \eg quotes. Each label should be on a new line. All other text in the \code{*.info} file is ignored.

\subsection{More than one chain in a single figure}

\begin{figure}[h]
\centering
\includegraphics[width=0.5\textwidth]{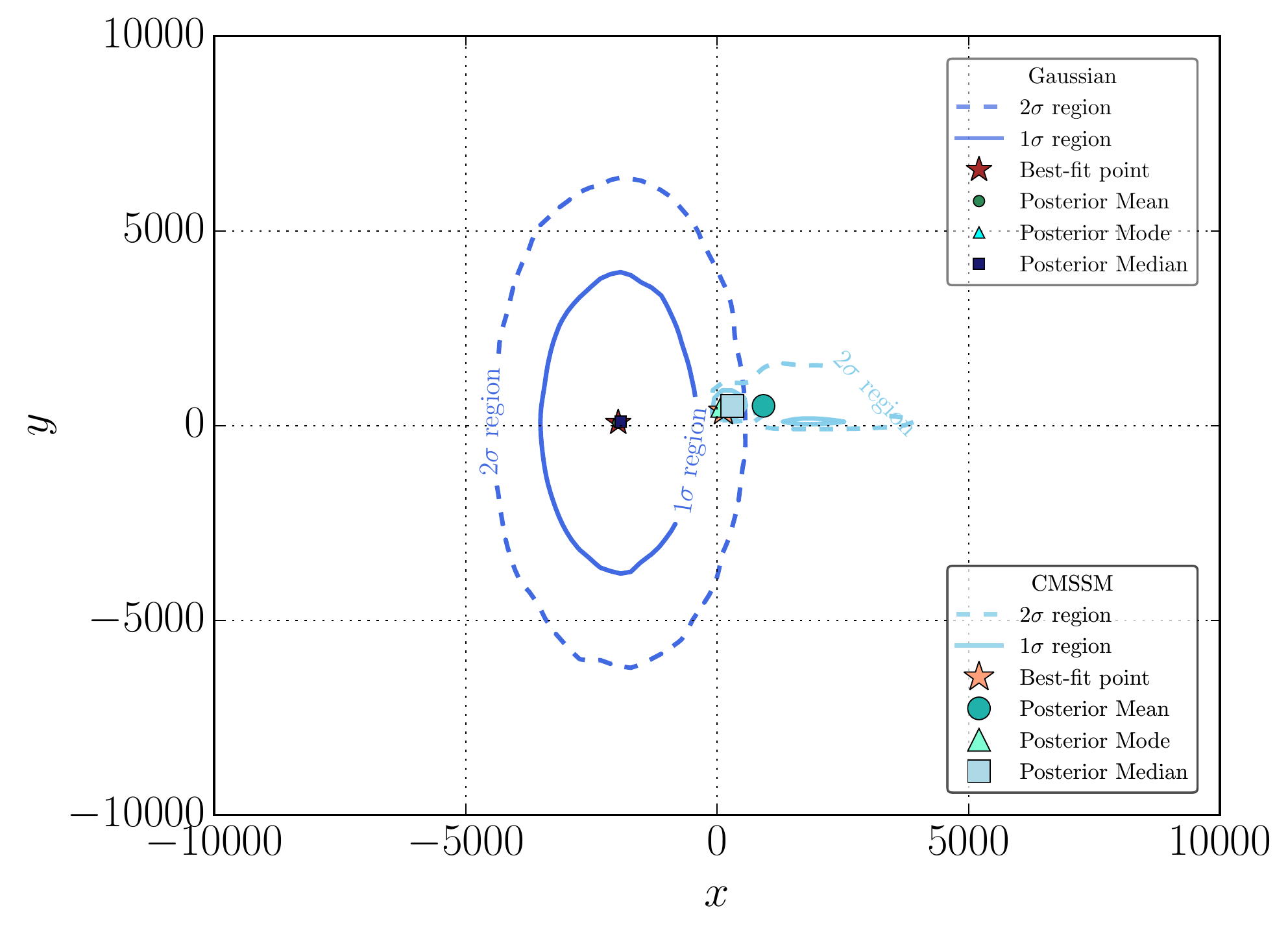}
\caption{Example output from \code{/example/two\_plots.py}, which uses \superplot as a library to plot statistics from two chains in a single figure.}
\label{fig:combined}
\end{figure}

We provide two example programs for plotting statistics from more than one chain in a single figure, \eg the posterior pdf with and without particular experimental data. The first, \code{/example/merge\_pkl.py}, overlays two or more pickled figure objects created in the \superplot GUI. For usage, see
\begin{lstlisting}
python merge_pkl.py --help  # Linux/Windows/Mac OSX
\end{lstlisting}
The second, \code{/example/two\_plots.py}, is a script that uses \superplot as a library to build a specialised plot. By default, it plots credible regions from the \code{/example/gaussian\_.txt} and \code{/example/SB\_MO\_log\_allpost.txt} chains, resulting in \reffig{fig:combined}.

\section{Summary and bug reporting}

We detailed a new application, \superplot, for plotting a chain from \eg \multinest. The application should be easy to install and use, and simplify the final step of parameter inference from a chain. To keep track of developments, report any bugs or ask for help, please see \url{https://github.com/michaelhb/superplot/issues}.

\appendix

\section{Statistical functions}\label{app:stat}

All statistical quantities in \superplot are calculated numerically. With the exception of the best-fit point, an error of about half a bin width is introduced in all quantities by binning. We advise that a user chooses the number of bins carefully, compromising between unnecessarily discarding information by coarse binning and noisy, comb-like distributions.  Out of the box, \superplot includes options for plotting many Bayesian and frequentist statistical quantities. By default, confidence intervals and credible regions are calculated at $1\sigma$ and $2\sigma$ two-tail significances, \ie for probabilities
\begin{equation}
\beta = 1 - \alpha = 2 \Phi(z) - 1
\end{equation}
for $z=1$ and $z=2$, where $\Phi$ is the cdf for the standard normal distribution. This results in $\alpha\approx0.32$ and $\alpha\approx0.046$. All statistical functions were tested against data drawn from Gaussian distributions with known properties.

\subsection{Bayesian}
We briefly describe the Bayesian statistical quantities in \superplot, including our conventions and ordering rules. See \eg \refcite{Gregory} for further discussion. Typically, quantities based upon the posterior pdf are not parameterisation invariant. We note only exceptions to this rule.

\begin{description}
\item[Marginalised posterior pdf] Defined
\begin{align}
p(x) &= \int p(x, y, z, \cdots) \dif y \dif z \dif \cdots\\
p(x, y) &= \int p(x, y, z, \cdots) \dif z \dif \cdots,
\end{align}
in the one- and two-dimensional cases, respectively. The pdf is calculated with a weighted histogram of samples in a chain with user-defined binning or by KDE as described in \refapp{app:KDE}.

\item[One-dimensional credible regions] Defined by a symmetric ordering rule, \ie an equal probability is contained in each tail:
\begin{align}
\int_{-\infty}^{x_a}  p(x) \dif{x} &= \int_{x_b}^{\infty} p(x) \dif{x} = \frac12 \alpha,\\
\int_{x_a}^{x_b} p(x) \dif{x} &= 1 - \alpha,
\end{align}
with a user-defined probability $\alpha$. This is invariant for monotonic transformations. One-dimensional credible regions are calculated by summing one-dimensional marginalised pdf and basic linear interpolation or by KDE as described in \refapp{app:KDE}.

\item[Two-dimensional credible regions] Defined by an ordering rule which is such that the credible region is the smallest region containing a given fraction of the total probability, \ie a credible region is the region $R$ such that $\int_{R}\dif{x} \dif{y}$ is minimised subject to the constraint that
\begin{equation}
\int_{R} p(x,y) \dif{x} \dif{y} = 1 - \alpha
\end{equation}
with a user-defined probability $\alpha$. Two-dimensional credible regions are calculated by finding the posterior density $p_\text{crit}$ such that
\begin{equation} 
\int\limits_{p(x,y) \ge p_\text{crit}} p(x,y) \dif{x} \dif{y} = 1 - \alpha
\end{equation}
is satisfied.

\item[Posterior mean] Defined
\begin{equation}
\langle x \rangle = \int x \cdot p(x) \dif{x}.
\end{equation}

\item[Posterior median] Defined such that
\begin{equation}
\int^{x^\prime}_{-\infty} p(x) \dif{x} = \int_{x^\prime}^\infty p(x) \dif{x} = 0.5.
\end{equation}
This is invariant for monotonic transformations. The posterior median is calculated in a similar manner to a one-dimensional credible region.

\item[Posterior modes] Parameter values (\ie bin centres in marginalised pdf) such that $p(x)$ or $p(x,y)$ is maximised. Note that the modes of $p(x)$ and $p(y)$ may not coincide with the mode of $p(x,y)$.
\end{description}

\subsection{Frequentist}
We briefly describe the frequentist statistical quantities in \superplot. Quantities based upon the profile likelihood are parameterisation invariant. See \eg \refcite{James:2006zz} for further discussion.
\begin{description}
\item[Profile likelihood] Defined
\begin{align}
\like(x) = \max_{y, z, \dots} \like(x, y, z, \cdots)\\
\like(x, y) = \max_{z, \dots} \like(x, y, z, \cdots),
\end{align}
in the one- and two-dimensional cases, respectively. Profile likelihood is calculated by binning the chain and finding the greatest likelihood within each bin.

\item[One-dimensional $\Delta \chi^2$] Defined
\begin{equation}
\Delta \chi^2(x) \equiv -2\ln \frac{\like(x)}{\max \like(x, y, z, \cdots)}.
\end{equation}
This may be chi-squared distributed by Wilks' theorem\cite{10.2307/2957648}.

\item[Confidence intervals] Defined via Wilks' theorem in the one-dimensional case as the region such that
\begin{equation}
-2\ln \frac{\like(x)}{\max \like(x, y, z, \cdots)} \le F_{\chi^2_1}^{-1}(1 - \alpha)
\end{equation}
and in the two-dimensional case, as the region such that
\begin{equation}
-2\ln \frac{\like(x,y)}{\max \like(x, y, z, \cdots)} \le F_{\chi^2_2}^{-1}(1 - \alpha)
\end{equation}
where $F_{\chi^2_n}^{-1}$ is the inverse cdf for a chi-squared distribution with $n$ degrees of freedom and $\alpha$ is a user-defined confidence level. We calculate whether each bin is excluded. Confidence intervals are not contiguous.

\item[Best-fit point] Defined as the parameter value (\ie not the bin centre) for the sample for which the likelihood is maximised.
\end{description}

\section{Kernel density estimation}\label{app:KDE}

There are drawbacks to estimating an unknown pdf by histogramming samples. Histogramming typically results in a jagged distribution that may be aesthetically unappealing as well as a poor estimate of the unknown pdf. One must, furthermore, judiciously pick the number of bins and the bin limits, both of which could affect the reconstructed pdf.

Kernel density estimation (KDE) is an alternative to histogramming \see{thompson1990nonparametric}. One replaces every sample with a kernel function and estimates the pdf by a weighted sum of the kernels over all samples:
\begin{equation}\label{eq:kde}
p(y) = \sum_i w_i K\left(\frac{y - x_i}{h}\right) \quad\text{where}\quad \int K\left(\frac{y - x_i}{h}\right) \dif{y} = 1 \quad\text{and}\quad \sum_i w_i = 1.
\end{equation}
The function $K$ is a kernel function and $h$ is the so-called bandwidth. \superplot includes only a Gaussian kernel function in one- or two-dimensions, with a variance or covariance matrix calculated from the samples. 

\begin{figure}
\centering
\hspace*{2.5mm}%
\subfloat[Weighted histogram with 30 bins per dimension]{\label{fig:1D_no_KDE}
\includegraphics[scale=0.25]{1D.pdf}
}%
\hspace*{4.5mm}%
\subfloat[Gaussian kernels and Scott's rule.]{\label{fig:1D_KDE}
\includegraphics[scale=0.25]{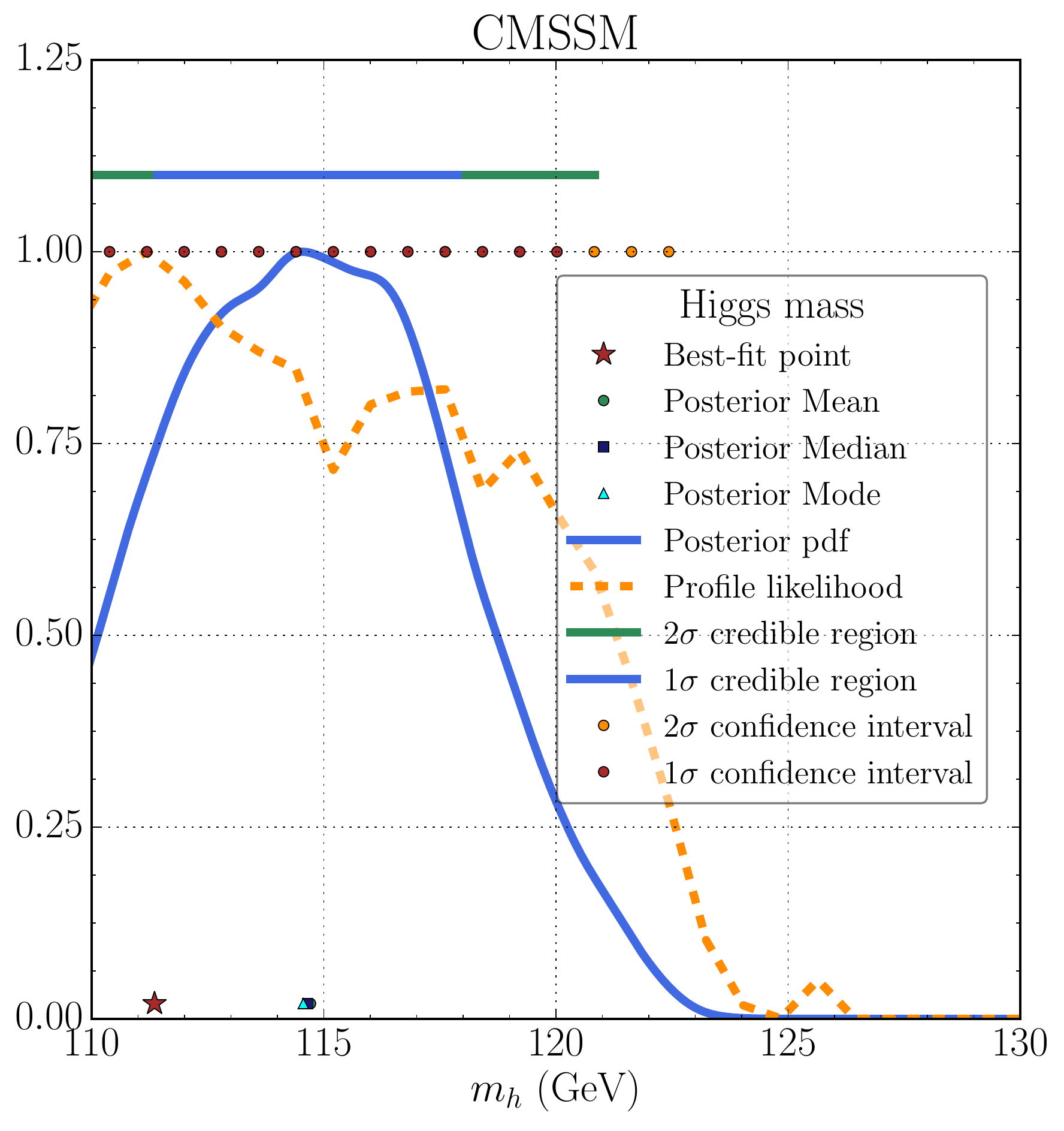}
}%

\subfloat[Weighted histogram with 30 bins per dimension]{\label{fig:2D_no_KDE_contour}
\includegraphics[scale=0.25]{2D.pdf}
}%
\subfloat[Gaussian kernels and Scott's rule.]{\label{fig:2D_KDE_contour}
\includegraphics[scale=0.25]{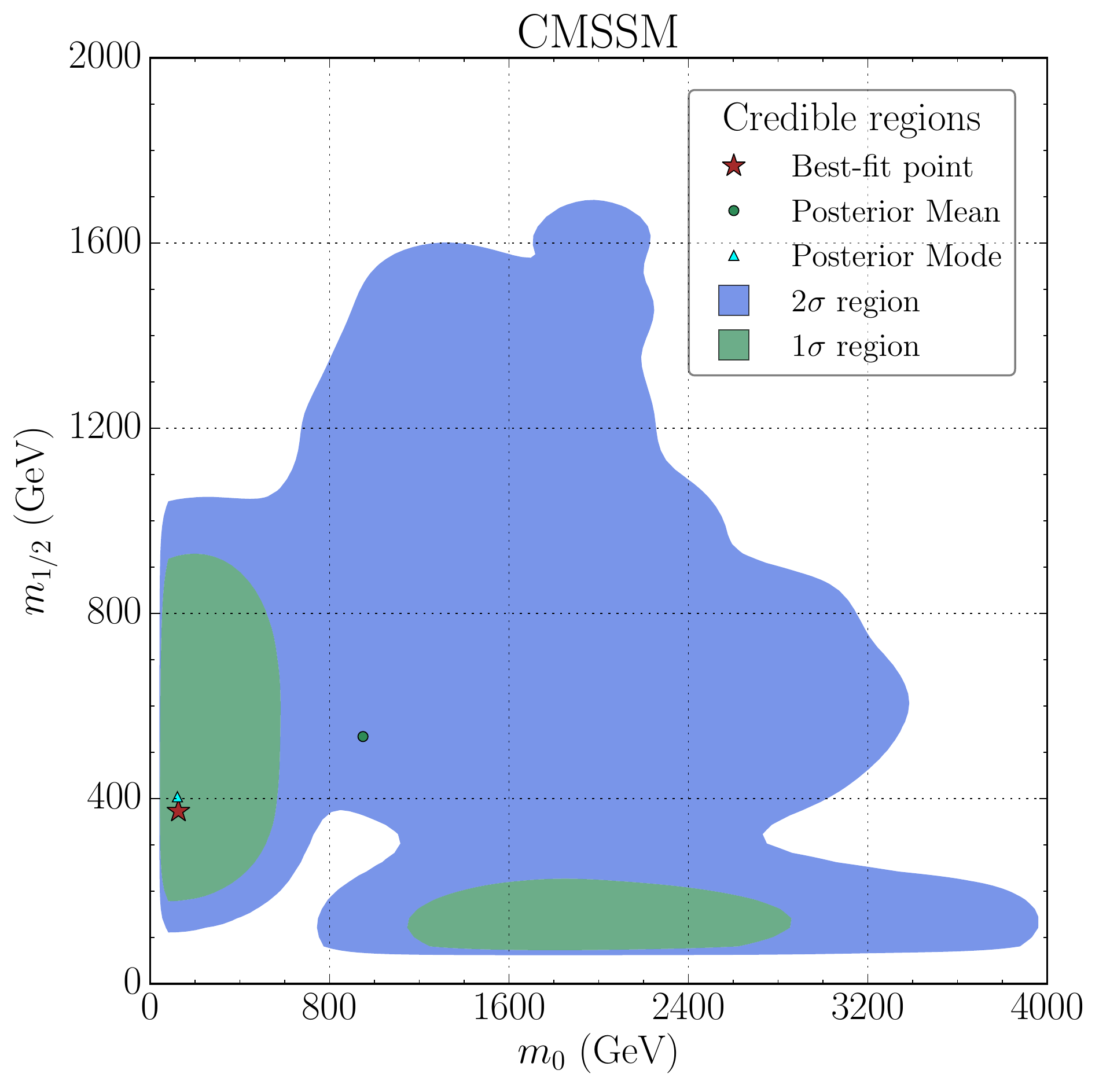}
}%

\subfloat[Weighted histogram with 30 bins per dimension]{\label{fig:2D_no_KDE_density}
\includegraphics[scale=0.25]{2D_density.pdf}
}%
\subfloat[Gaussian kernels and Scott's rule.]{\label{fig:2D_KDE_density}
\includegraphics[scale=0.25]{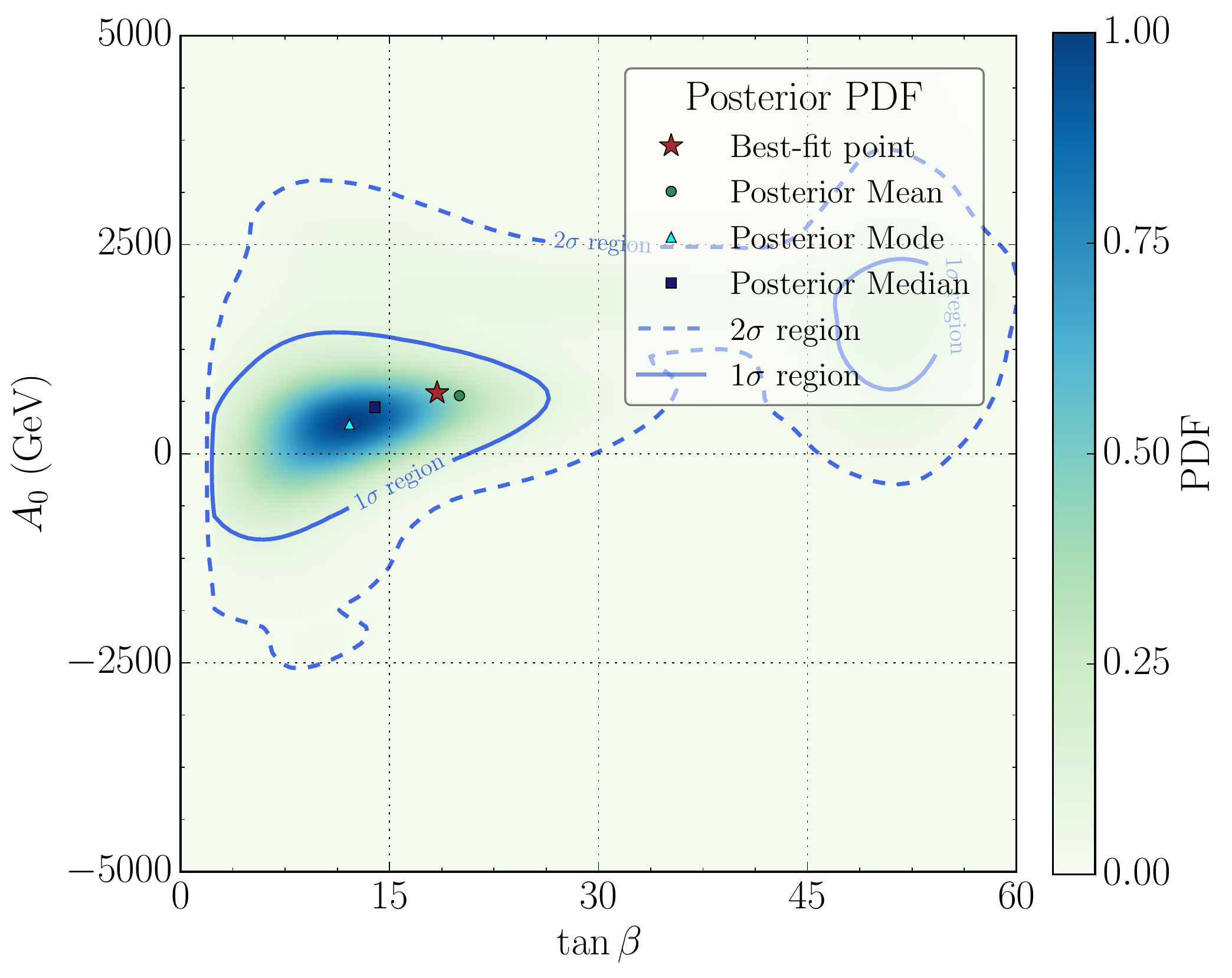}
}%
\caption{Weighted histograms with 30 bins per dimension contrasted with Gaussian kernel density estimation (KDE) with Scott's rule of thumb. All plots were produced by \superplot from a publicly released chain from \code{SuperBayeS}\cite{deAustri:2006jwj,sbweb}.}
\label{fig:kde}
\end{figure}

The bandwidth, which multiplies the standard deviation of Gaussian kernels, significantly impacts the estimate of the pdf. Indeed, picking the bandwidth is crucial: it could over-smooth, destroying genuine features of the unknown pdf, or under-smooth, and retain spurious noise in the estimate of the pdf. By default, \superplot uses Scott's rule of thumb\cite{scott2015multivariate} for the bandwidth,
\begin{equation}
h_\text{Scott} = n_\text{eff}^{-1/(d+4)} \quad\text{where}\quad n_\text{eff} = \frac{1}{\sum_i w_i^2},
\end{equation}
for $d$-dimensional KDE with an effective sample number $n_\text{eff}$. This is optimal for Gaussian distributions but may perform poorly for non-Gaussian, multi-modal distributions. Furthermore, whilst we insure that the KDE cannot over-spill outside a user-specified range (the bin range), there is no special treatment of boundaries, which may result in underestimates at the edges of unphysical regions. The bandwidth can be altered via the \code{bw\_method} variable in the \code{yaml} file \code{config.yml}, as described in \refsec{sec:custom}. Scott's rule (\code{scott}), Silverman's rule\cite{silverman1986density} (\code{silverman}),
\begin{equation}
h_\text{Silverman} = \left[\left(\frac{d+2}{4}\right) n_\text{eff}\right]^{-1/(d+4)} \approx 
\begin{cases} 
1.06 \cdot h_\text{Scott} & d=1\\
h_\text{Scott} & d=2
\end{cases},
\end{equation}
and user-defined bandwidths (\eg \code{0.01}) are supported. 

KDE is implemented via a modified version of \code{scipy.gaussian\_kde}.\footnote{The modifications support weighted samples and utilise discrete Fourier transforms. Development began from \refcite{stack}.} We finely bin samples in evenly spaced bins such that \refeq{eq:kde} becomes a discrete convolution,
\begin{equation}
p(y) = \sum_b w(x_b) K\left(\frac{y - x_b}{h}\right) = w \convolve K,
\end{equation}
where we sum over bins $b$ and $\cdot\convolve\cdot$ denotes a discrete convolution.
We evaluate the convolution by the convolution theorem,
\begin{equation}
w \convolve K = \idtft{\dtft{w} \cdot \dtft{K}},
\end{equation}
where $\dtft{\cdot}$ applies a discrete-time Fourier transformation (DTFT).
This well-known trick \see{arfken2012mathematical} reduces the complexity of the calculation, making it tractable for large numbers of samples, and is implemented in \code{scipy.signal.fftconvolve}.

To illustrate the pros and cons, in \reffig{fig:kde} we plot posterior pdf with and without KDE. For a simple one-dimensional pdf in \reffig{fig:1D_no_KDE} and \reffig{fig:1D_KDE}, KDE appears successful; the distribution is smoothed without distortion. The two-dimensional pdf in \reffig{fig:2D_no_KDE_contour} and \reffig{fig:2D_KDE_contour}, however, is problematic: genuine small features, such as the $1\sigma$ credible regions, appear over-smoothed and exaggerated in \reffig{fig:2D_KDE_contour} and the boundaries at $m_0=0$ and $m_{1/2} = 0$ appear over-smoothed. On the other hand, spurious small features, such as the noise in the $2\sigma$ credible region appear to be correctly smoothed away. In \reffig{fig:2D_no_KDE_density} and \reffig{fig:2D_KDE_density}, however, the pdf from KDE appears satisfactory as the contours are smoothed without obliterating or exaggerating genuine details. We advise that KDE is used with care, and disable it by default. 

\end{document}